\documentclass[showpacs,twocolumn,aps,epsfig,graphicx,floatfix]{revtex4}

\usepackage[dvips]{graphicx}
\newcommand{\be}{\begin{equation}}
\newcommand{\ee}{\end{equation}}

\newcommand{\bea}{\begin{eqnarray}}
\newcommand{\eea}{\end{eqnarray}}
\newcommand{\bd}{\begin{displaymath}}
\newcommand{\ed}{\end{displaymath}}
\newcommand{\bi}{\begin{itemize}}
\newcommand{\ei}{\end{itemize}}
\newcommand{\bc}{\begin{center}}
\newcommand{\ec}{\end{center}}
\newcommand{\bfl}{\begin{flushleft}}
\newcommand{\efl}{\end{flushleft}}
\newcommand{\bfr}{\begin{flushright}}
\newcommand{\efr}{\end{flushright}}
\newcommand{\f}{\frac}

\def\ra{\rightarrow}
\def\6{\partial}

  \def\D{\Delta}

\def\={\!\!\!&=&\!\!\!}
\def\+{\!\!\!&&\!\!\!+~}
\def\-{\!\!\!&&\!\!\!-~}

\begin{document}

\title{Evidence for a metallic--like state in the $T=0$ K phase
diagram of a high temperature superconductor}

\author{A.\ A.\ Schmidt}
\affiliation{Departamento de
Matem\'atica, CCNE, Universidade Federal de Santa Maria, 97105-900
Santa Maria, RS, Brazil}

\author{J.\ J.\ Rodr\'{\i}guez-N\'u\~nez}
\affiliation{Departamento de F\'{\i}sica--FACYT, Universidad de
Carabobo, Valencia 2001, Edo.\ Carabobo, Venezuela}

\author{ I. \c{T}ifrea}
\affiliation{Department of Physics and Astronomy, University of
Iowa, Iowa City, IA 52242, USA} \affiliation{Department of
Theoretical Physics, University of Cluj, Cluj 3400, Romania}

\date{\today}

\begin{abstract}
We examine the effects of a phenomenological pseudogap on the
$T=0$ K phase diagram of a high temperature superconductor within
a self-consistent model which exhibits a $d$-wave pairing
symmetry. At the mean-field level the presence of a pseudogap in
the normal phase of the high temperature superconductor is proved
to be essential for the existence of a metallic--like state in the
density versus interaction phase diagram. In the small density
limit, at high attractive interaction, bosonic--like degrees of
freedom are likely to emerge. Our result should be relevant for
underdoped high temperature superconductors, where there is a
strong evidence for the presence of a pseudogap in the excitation
spectrum of the normal state quasiparticles.

\end{abstract}


\maketitle

\section{Introduction}

One of the latest addition to the already very reach phase diagram
of high temperature superconductors (HTSC) is represented by the
so called pseudogap line ($T=T^*$). However, such a line is not
well defined, and it should be understood more as a boundary limit
which divides HTSC's temperature--density phase diagram into two
distinct regions \cite{timusk}. At temperatures $T_c<T<T^*$ ($T_c$
represents the critical temperature for the
normal--superconducting phase transition) and low doping, HTSC
display a gap in the energy spectrum of the single-particle
excitations associated to the normal state. Usually addressed as
the {\em pseudogap}, this anomalous property of HTSC was
originally seen in direct measurements of the single particle
excitation spectra in angle resolved photoemission spectroscopy
(ARPES) \cite{damascelli} and tunneling experiments \cite{renner}.
The origin of the pseudogap in HTSC above the transition critical
temperature $T_c$ is not clear, as it is not clear its connection
to the superconductivity phenomena. However, the symmetry of the
pseudogap follows the superconducting gap symmetry and is of
$d$-wave type \cite{damascelli}. The characteristic temperature
$T^*$ decreases with increasing the doping, eventually
intersecting the $T=0$ line at a {\em universal} critical point
\cite{loram}. The existence of such a critical point in the
cuprates phase diagram is still under debate, different approaches
relying on the fact that the $T^*$ line merges with the $T_c$ line
at a doping value of the order of the optimal doping, where the
transition temperature $T_c$ has its maximum \cite{timusk}.
Different authors \cite{randeria,giancarlo1,levin1} suggest that
the pseudogap and superconductivity in HTSC have the same cause,
the idea of preformed pairs being of main importance to these
theories. Other theories relate the pseudogap to antiferromagnetic
fluctuations \cite {schmalian}, spin-charge separation \cite{lee},
or incommensurate charge--density--wave instabilities
\cite{castellani}.

Experimental data for both normal and superconducting state in
HTSC prove that standard theories such as the Fermi liquid theory
of metals and BCS theory of superconductivity fail to correctly
describe the physics of these materials. Such a behavior is
attributed to the fact that correlations in HTSC are much stronger
than in standard superconductors. In particular, the small
coherence length of the Cooper pairs measured in HTSC (typically
about 10--20 \AA) suggested that HTSC belong to a class of
compounds which can be viewed in a crossover regime between BCS
superconductivity and Bose-Einstein condensation (BEC). From the
theoretical point of view, the possibility of a crossover
interpolation between BCS and BEC behavior was suggested previous
to the HTSC discovery by Eagles \cite{eagles} and latter on
developed by Leggett \cite{leggett}, and Nozi\'eres and
Schmitt--Rink \cite{nozieres}. More recently, various authors
\cite{randeria2} discussed the HTSC phase diagram in terms of a
BCS--BEC crossover, with the main conclusion that at least for the
underdoped region of the phase diagram a BEC--like description is
more appropriate to account for their unusual properties. On the
other hand one should mention that ARPES data prove that below
$T^*$ the pseudogap opens only around the $M$ points in the
Brillouin zone, other regions of the Fermi surface being gapless.
To account for such a behavior, one should consider a model with
strong correlations around the $M$ points and a weak interaction
for the rest of the Brillouin zone, in other words to consider a
two--gap model with a strong anisotropy \cite{perali}.

Here we investigate the effects of the pseudogap on the $T=0$ HTSC
phase diagram. We choose to treat the pseudogap at the
phenomenological level in order to understand its influence on the
physical properties of the superconducting state in HTSC. We
consider that both the pseudogap and the superconducting gap are
of $d$-wave symmetry and we neglect effects related to the
anisotropy. First we will analyze the role and the effects of the
pseudogap in the weak coupling limit, based on a modified BCS-like
model. The main result will be that the presence of the pseudogap
induces a critical value for the attractive electron--electron
interaction responsible for the formation of a superconducting
phase. Secondly, we extend this model to the strong interaction
limit, using an approach similar to the crossover interpolation
between BCS-like superconductivity  and BEC. In this limit we will
prove the existence of a metallic-like region in the phase diagram
oh cuprates, a region induced by the pseudogap presence.

\section{The Model}

The presence of the pseudogap in the energy spectrum of the normal
state electrons in HTSC materials will strongly influence their
physical properties not only in the normal state but also in the
superconducting state. To account for such a change, we will treat
the pseudogap at a phenomenological level. Accordingly, we
consider that the normal state can be described by a modified
one-particle Green's function which includes effects related to
the pseudogap presence \cite{tifrea},
\be\label{GreenFunction}
G^{-1}(\vec{k},\omega) =\omega+ \mu - \epsilon(\vec{k}) -
\Sigma(\vec{k},\omega)\;,
\ee
where $\omega$ represents the frequency and $\mu$ the chemical
potential. For the case of two--dimensional (2D) structures the
electronic energy dispersion is given by $\epsilon(\vec{k}) =
-2t[\cos(k_x)+\cos(k_y)]+4t\alpha'\!\cos(k_x)\cos(k_y)$, where
nearest--, and next--nearest--neighbor interactions are included.
The effects related to the pseudogap presence are introduced in
the self-energy, which we consider to be
\be\label{SelfEnergy}
\Sigma(\vec{k},\omega)=-E^{2}_g\phi^{2}(\vec{k})\,G_0(\vec{k},\omega)\;,
\ee
where $E_g$ is a constant, $\phi(\vec{k})=\cos{(k_x)}-\cos{(k_y)}$
is related to the $d$-wave pseudogap symmetry, and
$G_0(\vec{k},\omega)$ is the free electron Green's function. A
similar form of the electronic self-energy was used by Randeria
\cite{randeria} to explain the ARPES experimental results. We
assume that such a form of the self--energy will be valid
regardless the value of the coupling between the constituent
electrons. The $T=0$ K mean-field equation describing the
superconducting state for the case of a fully separable attractive
interaction in the presence of the pseudogap is obtained following
the standard procedure \cite{gorkov}

\be
\label{gap} \frac{1}{V}=
\frac{1}{\pi^{2}}\int^{\pi}_0\!\!\!dk_x\int^{\pi}_0\!\!\!dk_y\,
\frac{\phi^{2}(\vec{k})\,[\psi^+(\vec{k})-\psi^-(\vec{k})]}
{2\Delta_0\sqrt{\Delta^{2}_0+4E^{2}_g}}\;,
\ee
where
\be
\psi^{\pm}(\vec{k}) =
\f{A^{\pm}_0}{\sqrt{[\epsilon(\vec{k})-\mu]^{2} +
A^{\pm}_0\phi^{2}(\vec{k})}}\;,
\ee
\be
A^{\pm}_0 = E^{2}_g+\frac{1}{2}\, \left[\Delta^{2}_0 \pm
\Delta_0\sqrt{\Delta^{2}_0+4E^{2}_g}\, \right]\;,
\ee
and $\Delta_0$ represents the superconducting gap. In the absence
of the pseudogap ($E_g=0$) Eq.\ (\ref{gap}) reduces to the
standard BCS form. To include effects beyond the weak coupling
limit, along with Eq.\ (\ref{gap}) we consider also the $T=0$
density equation
\be\label{density}
\rho = \frac{1}{\pi^{2}}\int^{\pi}_0 dk_x\int^{\pi}_0
dk_y\left[\alpha_2(\vec{k}) + \alpha_4(\vec{k})\right]\;,
\ee
where
\begin{eqnarray}
\alpha_2(\vec{k})&\!\!=\!\!&
\frac{-[\omega^{2}_+(\vec{k})-(\epsilon(\vec{k})-\mu)^{2}]\,\,\,\,
G^{-1}(\vec{k},-\omega_2(\vec{k}))}{[\omega_2(\vec{k}) -
\omega_1(\vec{k})]\,[\omega_2(\vec{k}) - \omega_3(\vec{k})]\,
[\omega_2(\vec{k}) -
\omega_4(\vec{k})]}\;,\nonumber\\
\alpha_4(\vec{k})&\!\!=\!\!&
\frac{-[\omega^{2}_-(\vec{k})-(\epsilon(\vec{k})-\mu)^{2}]\,\,\,\,
G^{-1}(\vec{k},-\omega_4(\vec{k}))}{[\omega_4(\vec{k}) -
\omega_1(\vec{k})]\,[\omega_4(\vec{k}) - \omega_2(\vec{k})]\,
[\omega_4(\vec{k}) - \omega_3(\vec{k})]}\;, \nonumber
\end{eqnarray}
with
\bd
\omega^{2}_{\pm}(\vec{k}) = [\epsilon(\vec{k})-\mu]^{2}
+[\Delta^{2}_0/2+A^{\pm}_0]\,\phi^{2}(\vec{k})\;,
\ed
$\omega_1(\vec{k}) = |\omega_+(\vec{k})|$, $\omega_2(\vec{k}) = -
|\omega_+(\vec{k})|$, $\omega_3(\vec{k}) = |\omega_-(\vec{k})|$,
and $\omega_4(\vec{k}) = - |\omega_-(\vec{k})|$. The two
self-consistent Eqs.\ (\ref{gap}) and (\ref{density}) will be
solved together to include the strong interaction effects on the
chemical potential, a procedure which will allow us to interpolate
between weak and strong coupling limits.

\section{The pseudogap influence on the BCS--BEC crossover}

The possibility of a BCS--BEC crossover, namely an interpolation
between weak and strong attractive interaction between the
electrons, can be treated if we solve self--consistently Eqs.\
(\ref{gap}) and (\ref{density}). In the weak coupling limit (BCS
case), the value of the chemical potential is fixed at the Fermi
level, a self-consistent treatment of Eqs.\ (\ref{gap}) and
(\ref{density}) being irrelevant. However, things are completely
different in the strong coupling limit (BEC case), were due to a
strong interaction between the electrons we expect a dramatic
change in their chemical potential, and accordingly the two
equations have to be considered at the same time. In the following
we will discuss the solutions of the two coupled equations
\ref{gap}) and (\ref{density}), and possible effects due to the
pseudogap presence on the physical properties of HTSC as they
result from these solutions.

\begin{figure}[t]
\includegraphics[trim=40 135 40 135,clip,
                 width=.53\textwidth,angle=0]{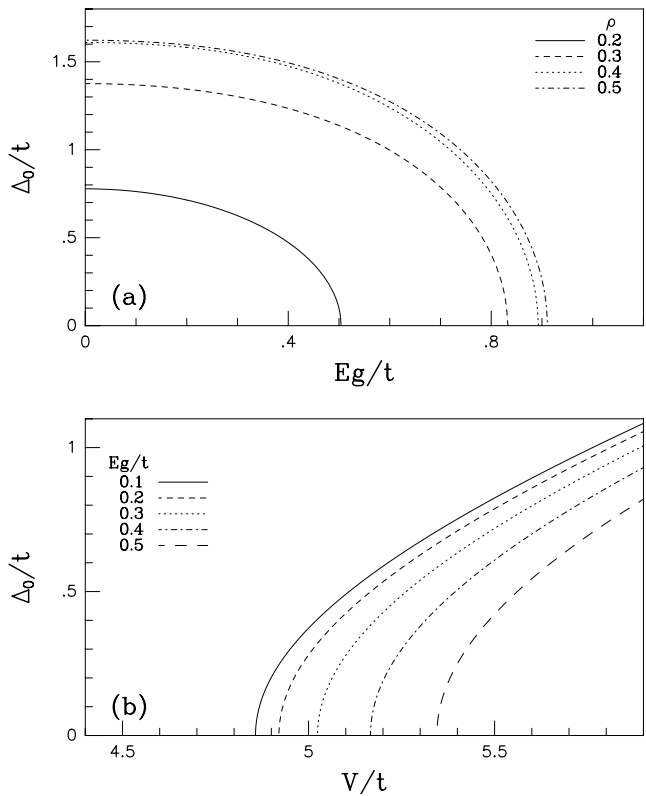}
\caption{a) The superconducting gap $\D_0/t$ as function of the
pseudogap parameter $E_g/t$ at different density values. b) The
superconducting gap $\D_0/t$ as function of the interaction
strength for different values of $E_g/t$ at a fixed density
$\rho=0.2$. (For both figures a $d$-wave model including next
nearest neighbors hopping was considered ($\alpha'=0.2$)).}
\label{fig1}
\end{figure}

\begin{figure}[t]
\includegraphics[trim=40 135 40 135,clip,
                 width=.53\textwidth,angle=0]{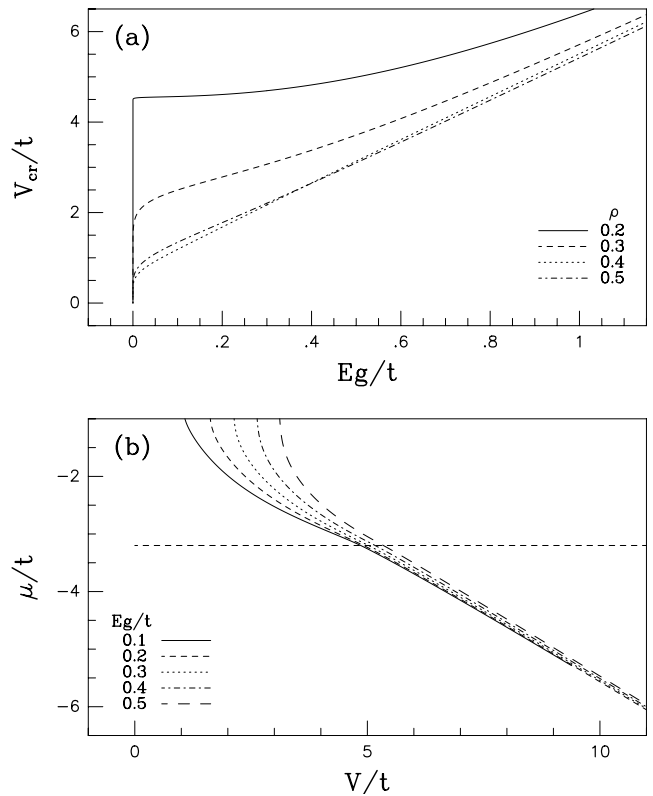}
\caption{a) The required critical interaction $V_{cr}$ as function
of the pseudogap parameter value $E_g$ at different density
values. b) The chemical potential as function of the interaction
strength $V$ for different values of the pseudogap parameter
$E_g$. The dashed horizontal line represents the minimum of the
conduction band, values for the chemical potential below this line
being a signal of bosonic-like degrees of freedom. (For both
figures a $d$-wave model including next nearest neighbors
interactions was considered ($\alpha'=0.2$)).} \label{fig2}
\end{figure}

The superconducting gap, $\D_0$, as it results from Eqs.\
(\ref{gap}) and (\ref{density}) is plotted in Fig.\ \ref{fig1}a as
function of the pseudogap value, $E_g$, for different values of
the system density. As the pseudogap value increases the
superconductivity gap decreases, the superconducting phase being
destroyed, a result which was already reported in Ref.
\cite{tifrea2}.
For $E_g=0$, the gap equation (\ref{gap}) reduces to the standard
BCS gap equation.  For the BCS case the gap equation has a real
solution for any value of the attractive interaction $V$, the
presence of the superconducting state being signaled by a nonzero
value for the superconducting order parameter $\D_0$. When the
self-consistent problem is solved in the presence of a pseudogap,
a critical attractive interaction is required for the existence of
a superconducting state. Fig.\ \ref{fig1}b presents the
superconducting gap parameter $\D_0$ as function of the attractive
interaction $V$ for different values of the pseudogap parameter
$E_g$. The plot clearly identify a critical value for the
interaction required for a nonzero value of the gap parameter.
Moreover, the critical interaction value can be extracted as it
follows. Let us consider the limit situation in which $\Delta_0\ra
0$. In such circumstances, Eq.\ (\ref{gap}) becomes
\be\label{vcritic}
\f{1}{V_{cr}}=\f{1}{8\pi^2}\int_0^\pi dk_x\int_0^\pi dk_y\,
\f{\phi^2(\vec{k})\,E_g^2 + 2\,E^2(\vec{k})}
{[\phi^2(\vec{k})\,E_g^2+E^2(\vec{k})]^\frac{3}{2}}\;,
\ee
where $E(\vec{k}) = \epsilon(\vec{k})-\mu$. Fig.\ \ref{fig2}a presents
results for the critical interaction
$V_{cr}$ as function of the pseudogap parameter $E_g$ for
different values of the system density. The results were obtained
using a self-consistent calculation which involves both Eq.\
(\ref{vcritic}) and the density equation (\ref{density}). In the
weak coupling case, when the chemical potential is fixed at the
Fermi level the density equation becomes irrelevant, and the value
of the critical interaction $V_{cr}$ can be obtained analytically.
The corresponding solutions has the form
$1/V_{cr}=(m/4\pi)\ln{[0.892 \; W/(E_g)]}$, with $W$ being a
cutoff such that $W\gg E_g$. It is clear at this point that for
$E_g\ra 0$, $V_{cr}\ra 0$, which means that in the absence of a
pseudogap the superconducting state exists no matter how small the
attractive electron-electron interaction is. However, the presence
of a pseudogap requires a critical interaction for a nonzero
superconducting gap, meaning that if $V<V_{cr}$ a metallic--like
state will be present in the HTSC phase diagram in the weak
coupling regime.

Fig.\ \ref{fig2}b presents the chemical potential as function of
the attractive interaction strength $V$ for different values of
the pseudogap parameter $E_g$. The horizontal line at $\mu
/t=-3.2$ represents the bottom of the tight-binding band. It is
well know that in the weak coupling regime, the fermionic nature
of the electronic system is conserved despite the fact that
electrons form pairs. Such a behavior of the system is signaled by
a value of the chemical potential above the bottom of the
tight-binding band, equivalent with a positive chemical potential
for the continuous model. On the other hand, in the strong
coupling regime, at high attractive interactions, the value of the
chemical potential for all curves is situated bellow this minimum
of the tight-binding band, the system manifesting bosonic--like
degrees of freedom. In this limit the attraction leads to small
sized electronic pairs which behave like composite bosons able to
undergo a BEC. The situation becomes much more complicated in the
intermediate interaction regime, were the system should be
described by a crossover theory which interpolates between the
standard BCS and BEC pictures.

The density--interaction phase diagram of the $T=0$ HTSC usually
consists in two different regions, associated to fermionic and
bosonic degrees of freedom, respectively \cite{giancarlo2}. For
the case of a system with $s$-wave symmetry of the superconducting
order parameter, it was proved that a crossover between fermionic
and bosonic degrees of freedom is possible at any values of the
system density \cite{nozieres}. On the other hand, the situation
for the case of a $d$-wave symmetry of the superconducting order
parameter is much more controversial. den Hertog \cite{denHertog}
and Soares {\em et al}.\ \cite{jose1} claim that a non-pairing
region, metallic--like phase, is present in the phase diagram at
low attractive interactions. However, such a claim is in
contradiction with the standard $d$-wave solution of the weak
coupling BCS equation, which proves for the continuous model the
existence of a superconducting state regardless the value of the
attractive interaction between the electrons \cite{chubukov}.

\begin{figure}[t]
\includegraphics[trim=40 135 40 135,clip,
                 width=.53\textwidth,angle=0]{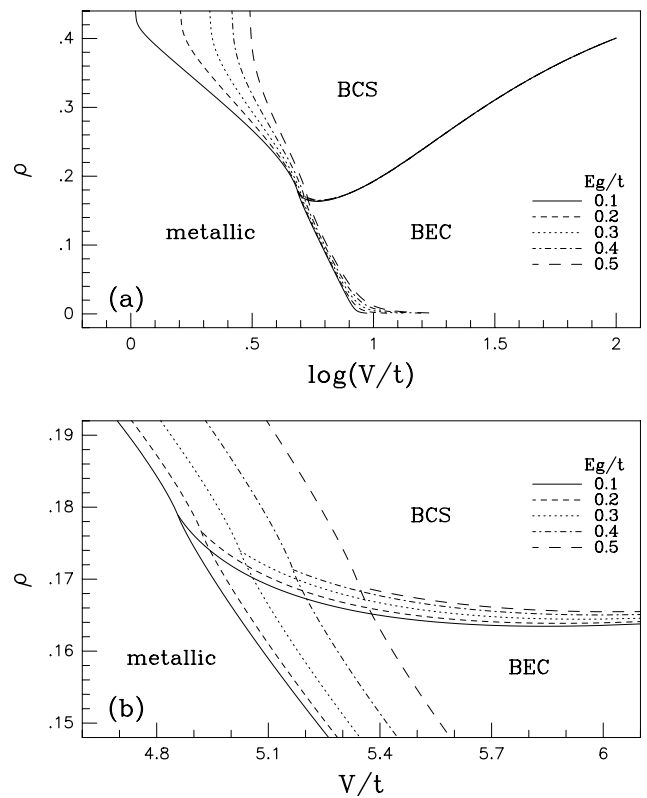}
\caption{a) Possible phase diagram ($\rho$ vs. $\log{(V/t)}$) for
a $d$-wave superconductor for different values of $E_g$.
Metallic-like ($\D_0/t=0$), BCS-like, and BE-like phases are
properly identified. b) A better insight of the differences
induced by different values of the pseudogap $E_g$. For both
figures a $d$-wave model including next nearest neighbors
interactions was considered ($\alpha'=0.2$).} \label{fig3}
\end{figure}

Fig.\ \ref{fig3} presents the effects of the pseudogap parameter
$E_g$ on the density--interaction phase diagram of the HTSC at
$T=0$. As we previously discussed, the consequences of the
pseudogap presence in the normal state of the HTSC are summarized
by the need of a critical attractive interaction in order to form
pairs which will be subject of a BCS or BEC--like phase.
Fig.\ \ref{fig3}a presents a possible phase diagram for HTSC
materials at different values of the pseudogap parameter $E_g$. In
the weak coupling limit, when the interaction between the
constituent electrons is small ($V<V_{cr}$), a metallic--like
state is present. Such a state is identified by the condition
$\D_0=0$, i.e., a superconducting state is absent. However,
calling this state metallic, may be very inappropriate, as the
presence of the pseudogap will have a strong effect on the system
properties, meaning that the standard Landau theory of metals may
not work in this situation. In the strong coupling regime
($V>V_{cr}$), the attractive interaction between the constituent
electrons may lead to the formation of electronic pairs which can
form a BCS or BEC--like state. The difference between the two
possible states is made by the value of the chemical potential,
which at the border between the two phases equals the energy value
of the minimum of the tight-binding band. As Fig.\ \ref{fig3} shows,
bosonic--like degrees of freedom are possible only at small
densities and high attractive interactions. This result is in
agreement with previous calculations in the literature
\cite{giancarlo1,levin1}. The behavior of the system looks similar
for different nonzero values of the pseudogap parameter. For any
value of the pseudogap $E_g$ all three phases of the phase
diagram, metallic, BCS, and BEC--like are present, (see
Fig.\ \ref{fig3}b). The phase diagram was obtained considering only
direct interactions between the constituent electrons of the
system. Once the electronic pairs are formed, especially in the
strong coupling limit, there is a possibility that pair--pair
interaction will be relevant to the problem, approximations beyond
the usual mean--field treatment being required \cite{popov}.

\section{Conclusions}

In conclusion we have obtained the phase diagram of a HTSC at
$T=0$ K in the case of a $d$-wave symmetry of the order parameter.
The role of the pseudogap parameter $E_g$ on the phase diagram was
considered in a phenomenological way, by introducing its effects
as corrections to the free electron state via an additional term
in the self-energy. The form of such a term can be justified by
the analysis of the experimental data obtained from ARPES
measurements \cite{randeria}. Our calculations, both analytical
and numerical, proved that the pseudogap effects can be drastic on
the density--interaction phase diagram of HTSC, a metallic--like
state being identified in the weak coupling regime. The question
that this state is pure metallic, or is of a different nature, is
still open, further calculations being underway. Our calculations
should be relevant for the underdoped cuprates, where the presence
of the pseudogap was experimentally proved. However, if the
presence of the pseudogap will be proved also in the overdoped
region of the HTSC temperature--doping phase diagram the
calculations presented here will correctly describe this region of
the phase diagram too.


\begin{acknowledgments}
\noindent We are very grateful to H.\ Beck, R.\ Micnas, R.\ Medina, I.\
Bonalde, S.\ G.\ Magalh\~aes and F.\ Kokubun for interesting
discussions. The numerical calculations were performed at LANA
(Departamento de Matem\'atica, UFSM). We thank
$C.D.C.H.$-$U.C.$--Venezuela (Project 2001-013) and
$FONACYT$--Venezuela for financial support (S1-2002000448). This
work was partially supported by the Brazilian agencies FAPERGS and
CNPq. One of the authors ($J.J.R.N.$) is a Fellow of the
Venezuelan Program of Scientific Research ($P.P.I.$--$IV$) and a
Visiting Scientist at $IVIC$--Venezuela.
\end{acknowledgments}

\end{document}